\documentclass[bibyear]{aa} 

\usepackage{graphicx}
\usepackage{amsmath}
\usepackage{float} 
\usepackage{caption}

\usepackage{txfonts}

\begin{document}

   \title{Exploratory scenarios for the differential nuclear and tidal evolution of TZ Fornacis}

   \subtitle{}

\author{    A. Claret~\inst{1,2}
 }

   \offprints{A. Claret, e-mail:claret@iaa.es}
\institute{Instituto de Astrof\'{\i}sica de Andaluc\'{\i}a, CSIC, Apartado 3004,
 18080 Granada, Spain
 \and
 Dept. F\'{\i}sica Te\'{o}rica y del Cosmos, Universidad de Granada, 
 Campus de Fuentenueva s/n,  10871, Granada, Spain\
}

   \date{Received; accepted; }

 \abstract
   {}
   {TZ Fornacis is a double-lined eclipsing binary system with  similar masses  (2.057$\pm$0.001 and 1.958$\pm$0.001 $M_{\odot}$)  
        but characterized by very different radii (8.28$\pm$0.22 and 3.94$\pm$0.17 $R_{\odot}$). This similarity in terms of mass   makes it possible  to study 
        the system's differential stellar evolution as well as some aspects of its tidal evolution. With regard to its orbital elements, it was 
        recently confirmed that its orbit is circular with an orbital period of 75.7 days. The   less massive component 
        rotates about 17 times faster than the primary one, which is synchronized with the mean orbital angular velocity. 
        Our main objective in this work is to study both the nuclear and the tidal evolution of the system.}
    {To model the TZ For system, we used the MESA package, computing  the grids using the exact observed masses, radii, 
        and effective temperatures as input, and then varying the metallicity, the core overshooting amount, and the mixing-length 
        parameter.  A $\chi^2$ statistic was used to infer the optimal values
        of the core overshooting and the mixing-length parameters. The same procedure was used to generate rotating 
        models with the GRANADA code. The respective errors in the average age of TZ For  were less than 5\%. On the 
        other hand,  the differential equations that govern the tidal evolution were integrated using the fifth-order 
        Runge-Kutta method, {\b with a tolerance}  
        of 1$\times$10$^{-7}$.
        }
   { We explored two scenarios regarding the initial eccentricities: a high one (0.30) and a case of an initial  
        circular orbit. A good agreement has been found between the observational values of the eccentricity, synchronism 
        levels, and orbital period with the values predicted by the integration of the tidal evolution equations. The influence 
        of the friction timescale on the evolution of the orbital elements of TZ For is also studied here. The orbital elements 
        most affected by the uncertainties in the friction timescale are the synchronism levels of the two components. On the 
        other hand, we used the properties of the rotating models generated by the GRANADA code as the initial angular velocities 
        instead of using trial values. In this case, comparisons between the theoretical values of the orbital elements and 
        their observed counterparts also lead to a good interagreement.}
   {}

   \titlerunning {TZ For tidal evolution }
   \maketitle
%

\section{Introduction}

TZ Fornacis (TZ For) is a very evolved  double-lined eclipsing binary (DLEBS) that  was  first accurately  investigated by Andersen et al. (1991); see also Torres et al. (2010). Furthermore,  TZ Gallene et al. (2016) revised the absolute dimensions of this system using  interferometric 
observations combined with new  radial velocity measurements and   confirmed that the system demonstrates very similar components in mass 
(2.057$\pm$0.001 and 1.958$\pm$0.001 $M_{\odot}$), but with very different radii: 8.28$\pm$0.22 and 3.94$\pm$0.17 $R_{\odot}$, respectively. 
The effective temperatures for both components are 4930$\pm$30 K and 6650$\pm$200 K and  the orbital period is $\approx$ 75.7 days. This 
mass similarity and the accuracy of the absolute dimensions makes TZ For a good laboratory for the study of its differential stellar evolution as 
well as some aspects of its tidal evolution. In fact,  both  components  evolve quite similarly in the main sequence, however,  during the 
very fast phases of their  evolution, we can significant changes in the effective temperatures 
and also in their radii, apsidal motion constants, and moment of inertia -- on the basis of stellar evolutionary models. Concerning its tidal evolution, Andersen et al. (1991) established 
a circular orbit for TZ For and these authors also found that  while the primary rotates synchronously with the mean orbital angular velocity, the secondary rotates 
approximately 16 times faster than its respective value of synchronisation, which is on the order of 2.6$\pm$0.1 km/s. These results are 
consistent  with those obtained by Gallene et al. (2016).

One of the first attempts to explain some aspects of both nuclear and tidal evolution was carried out by Claret \& Giménez (1995), who  
adopted the hydrodynamical mechanism proposed by Tassoul (1987). However, that study did not provide a clear scenario of the stellar and 
dynamic evolution of the system since,  at that time, the current position of the system in the HR diagram was not yet fully clear. 
Second, timescales for the circularisation and synchronisation were used  (assuming a constant  orbital period) instead of formally 
integrating the differential equations that govern the behavior of angular velocities, orbital period, and eccentricity. In addition, 
Rieutord (1992) presented some criticism  on the hydrodynamical 
mechanism introduced by Tassoul (1987) regarding the inefficiency in reducing
the synchronization time of the large-scale flows driven by
Ekman pumping in the spin-up and down of a tidally distorted star.  Later on, Tassoul \& Tassoul (1996) refuted this argument. For a more extensive  
discussion on this subject,  we refer to Rieutord (1992),  Tassoul \& Tassoul (1996), 
and Claret \& Cunha (1997). In fact, Claret \& Cunha (1997) found some disagreements between the tidal evolution theories 
predictions and the observational values: while the hydrodynamic mechanism was indeed shown to be too efficient in circularizing the orbits, 
the opposite happened with the tidal torque process. 

A number of years later,  Claret (2011) tried a new approach to the problem.  Instead of adopting the timescales as in Claret \& Giménez (1995), which are valid only for low eccentricities and small departures from 
the synchronism, a more rigorous treatment was carried out.  These authors found that while the first approximation can be considered roughly acceptable
at least for the present status of TZ For (but not necessarily in the past), the second  is not satisfied given the observed level of   
asynchronism of the secondary. To further improve on the previous method, the  differential equations that govern the 
tidal evolution (eccentricity, angular velocities, and orbital  period) were explicitly integrated adopting appropiate models computed  
with  the  GRANADA code (Claret 2004). Another point that was clarified in the 2011 paper was the position of the primary, which was located on the clump  during the helium-burning phase.  

In this paper, we propose a revision to the nuclear and dynamic  evolution of the system by taking into account the most recent 
determinations of the absolute dimensions by Gallene et al. (2016).  
 Here, the expression nuclear evolution refers to the changes in effective temperatures, radii, luminosities, moments of inertia, 
        and other effects caused by nuclear reactions in the stellar interior.  To model the TZ For system, we computed stellar
evolutionary tracks with the MESA Package,  adopting the methodologies described  in detail by Claret \& Torres (2016, 2017, 2018, 2019). 
In addition, an updated version of  the  GRANADA code was used to simulate stellar rotation based on the assumption of a  solid body configuration. 

The paper is structured as follows. In Sect. 2, we  briefly describe  the stellar evolution code MESA and our methodology for
deriving, for each star, the semi-empirical value of f$_{ov}$ (overshooting parameter, see below) by  comparing the observed absolute 
dimensions with a series of  grids of stellar models for both components of TZ For. In this section, we also introduce the 
differential equations of tidal evolution.  In Sect. 3, we analyse the stellar and tidal evolution for TZ For and summarise our results. 
Finally,   in Appendix A, we describe the method for generating rotating models introduced by  Kippenhahn \& Thomas (1970)  and implemented in the 
GRANADA code and  adapted to the particular case  of TZ For. 
 
\section {Stellar models and the differential equations of tidal evolution  }

\subsection {Stellar models}

The  stellar evolutionary tracks used to fit the absolute dimensions of TZ For were computed using  the Modules 
for Experiments in Stellar Astrophysics package (MESA; Paxton et al. 2011, 2013, 2015) version 7385.  
Microscopic diffusion was included  and mass loss  was considered folowing the formulation by  Reimers (1977), with
an efficiency coefficient of $\eta$=0.2 (not to be confused with $\eta$ of the Radau equation). 
For  the convective envelopes, we employed the  mixing-length formalism (Böhm-Vitense 1958), where $\alpha_{MLT}$ is a 
free parameter. The calibrated value  of the mixing-length parameter for the Sun in these models is 
$\alpha_{MLT}$ = 1.84 for Z$_{\odot}$ = 0.0134.  For the opacities, we adopted the mixture given 
by Asplund et al. (2009). The enrichment law used in pairings with such  opacities 
was $\Delta$Y/$\Delta$Z = 1.67, with the primordial helium Y$_p$  = 0.249 following Ade et al. (2016).  
Convective core overshooting was considered in the diffusive approximation which is  characterized by 
the free parameter f$_{ov}$.  For more details, we refer to Freytag et al. (1996) and Herwig et al. (1997).  The grids were computed 
starting from the pre-main sequence (PMS)  and the effects of rotation were ignored for this set of models. 

As previously commented, a series of papers by Claret \& Torres (2016, 2017, 2018, 2019) used stellar evolutionary 
models  and a select sample of 50 well-measured detached DLEBS to calibrate 
the dependency of core overshooting on stellar mass.
Here, we use  the results of one of these papers  (Claret \& Torres (2018)) as a reference where the case of TZ For had already been analysed. 
In that paper, several grids were computed for this system using as a main input  the exact  observed masses, radii, and effective 
temperatures varying the metallicity, as well as the core overshooting amount and the mixing-length parameter. In the search for the  best 
solution for the two components, we adopt a $\chi^2$  statistic to infer the optimal values 
of the overshooting  and the mixing-length parameters. Due to  intrinsic limitations in the stellar models (opacities, equations 
of state, mass loss, etc.) and considering the observational errors, the derived ages for the two components were allowed to
differ by up to 5\%.   The best fits for 
TZ For, adopting  the mixture by Asplund et al. (2009) and the enrichment law described above, are as follows: 
$\alpha_{MLT1}$ = 1.91, $\alpha_{MLT2}$ = 1.85 and f$_{ov1}$ = 0.017, and  f$_{ov2}$ = 0.015 for a metallicity 
of Z = 0.015. The derived mean age of the two components is 1.13$\pm$0.04 Gyears. Figure 1 shows the resulting 
HR diagram where the primary is located at the clump. We note that our solution is slightly different from that found by 
Gallene et al. (2016) since these authors located the secondary near the red hook. Another interesting aspect derived from 
Claret \& Torres (2018) is that if  a different 
enrichment law is adopted for TZ For, for instance, $\Delta$Y/$\Delta$Z = 1.0, we obtain essentially the same result:  
the primary is also in the core helium-burning phase (clump) and the secondary on the subgiant branch. This suggests that the f$_{ov}$ parameter is insensitive to helium content for this particular case. 

 On the other hand,  TZ For was also recently studied by Costa et al. (2019), who used a different method from that of Claret \& Torres. The basic difference  between the two methods is that Costa et al. (2019) introduced 
        rotation into their models (taking into account rotational mixing) by assuming constant values of core overshooting and MLT parameters. According to these authors, they found: a good agreement with models computed with
 fixed overshooting parameter, $\lambda_{ov}$ = 0.4, and initial rotational rates, $\omega$, uniformly distributed
a in a wide range between 0 and 0.8 times the break-up value, at varying initial mass. We also note that their definition of  
$\lambda_{ov}$ is different from the one we used (diffusive approximation, characterized by the parameter f$_{ov}$).

\begin{figure}
        \includegraphics[height=8.cm,width=6cm,angle=-90]{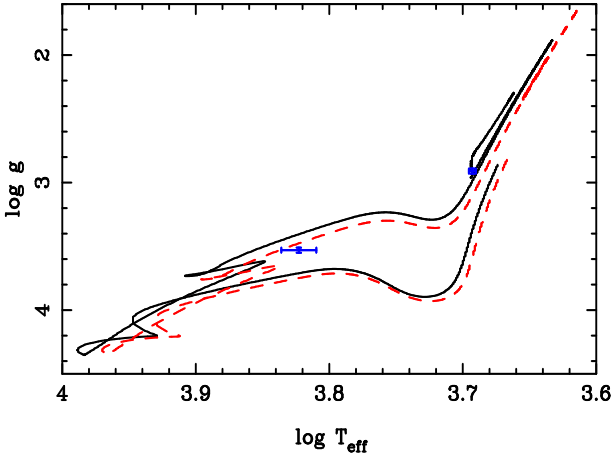}
        \caption{HR diagram for TZ For. The models were calculated adopting $Z = 0.015$. Solid line indicates the primary component while 
                the dashed one represents the secondary.}
\end{figure}

\subsection {Differential equations of tidal evolution}

Here, we adopt the differential equations for the tidal evolution in the frame of  the weak friction model (Hut 1981). 
The respective  equations for an initial  eccentricity different from zero can be written as:

\begin{eqnarray}
        {de\over {dt}} = -{27 k_{2 1}\over{t_{F1}}} q (q +1) \left(R_1\over{A}\right)^8 {e \over{(1-e^2)^{13/2}}} \nonumber \\
        \left( f_3 - 11/18(1-e^2)^{3/2} 
        f_4 {\Omega_1\over{\omega}}\right),
\end{eqnarray}

\begin{eqnarray}
        {dA\over {dt}} = -{6 k_{2 1}\over{t_{F1}}} q (q +1) \left(R_1\over{A}\right)^8 {A \over{(1-e^2)^{15/2}}}  \nonumber \\ \left( f_1 - (1-e^2)^{3/2} 
        f_2 {\Omega_1\over{\omega}}\right),
\end{eqnarray}

\begin{eqnarray}
        {d\Omega_1\over {dt}} = {3 k_{2 1}\over{t_{F1} \beta_1^2}} q^2 \left(R_1\over{A}\right)^6 {\omega \over{(1-e^2)^{6}}} 
        \left( f_2 - (1-e^2)^{3/2} f_5 {\Omega_1\over{\omega}}\right),
\end{eqnarray}

\begin{eqnarray}
        {d\Omega_2\over {dt}} = {3 k_{2 2}\over{t_{F2} \beta_2^2}} q_2^2 \left(R_2\over{A}\right)^6 {\omega \over{(1-e^2)^{6}}} 
        \left( f_2 - (1-e^2)^{3/2} f_5 {\Omega_2\over{\omega}}\right).
\end{eqnarray}

\noindent
In the above equations,  $e$ is the orbital eccentricity, A is the semimajor axis, $\Omega_{i}$ 
is the angular velocity of the component $i$, and $\omega$ is the mean orbital angular velocity, while  
$\beta_i$ is the radius of gyration of the component $i$, then k$_{2 i}$ is the apsidal motion 
constant of the component $i$ and R$_i$ is the radius of the component $i$, and $q = M_2/M_1$,   $q_2 = 
M_1/M_2$ and t$_{F}$ is an estimation of the timescale of tidal friction which is given  by: 

\begin{eqnarray}
        t_{F} = (MR^2/L)^{1/3},
\end{eqnarray}
\noindent
where L is the luminosity, R the radius, and M the stellar mass. For a solar-type star, t$_F$ $\approx$ 0.43 year.

The  functions f$_k$ can be written as :

\begin{eqnarray}
        f_1 = 1 + {31\over{2}} e^2+ {255\over{8}} e^4 + {185\over{16}} e^6 + {25\over{64}} e^8,
\end{eqnarray}

\begin{eqnarray}
        f_2 = 1 + {15\over{2}} e^2+ {45\over{8}} e^4 + {5\over{16}} e^6 
\end{eqnarray}

\begin{eqnarray}
        f_3 = 1 + {15\over{4}} e^2+ {15\over{8}} e^4 + {5\over{64}} e^6, 
\end{eqnarray}

\begin{eqnarray}
        f_4 = 1 + {3\over{2}} e^2+ {1\over{8}} e^4, 
\end{eqnarray}

\begin{eqnarray}
        f_5 = 1 + 3 e^2+ {3\over{8}} e^4.
\end{eqnarray}

For the case of  very small initial  eccentricities we have:

\begin{eqnarray}        
  {de\over{dt}} \rightarrow 0,
\end{eqnarray}  

\begin{eqnarray}        
          {dA\over {dt}} \approx  c_1(1-{\Omega_1\over{\omega})}),
\end{eqnarray}          
        
\begin{eqnarray}
  {d\Omega_1\over {dt}} \approx  c_2 (\Omega_1-\omega),
\end{eqnarray}  

\begin{eqnarray}
          {d\Omega_2\over {dt}} \approx  c_3 (\Omega_2-\omega),
\end{eqnarray}  

\noindent
where 
\begin{eqnarray}
 c_1=-{6 k_{2 1}\over{t_{F1}}} q (q +1) \left(R_1\over{A}\right)^8 A,
\end{eqnarray}

\begin{eqnarray}
  c_2= {3 k_{2 1}\over{t_{F1} \beta_1^2}} q^2 \left(R_1\over{A}\right)^6,
\end{eqnarray}

\begin{eqnarray}
 c_3={3 k_{2 2}\over{t_{F2} \beta_2^2}} q_2^2 \left(R_2\over{A}\right)^6.
\end{eqnarray}

On the other hand, the corresponding  theoretical internal structure constants, k$_j$, 
were computed as a function of time  for each component during its evolution  through the integration of 
the differential equations on a Radau order of $j$: 

\begin{eqnarray}
 {a {d\eta_{j}(a)\over da}}+ {6\rho(a)\over\overline\rho(a)}{(\eta_{j}\!+\!1)}+
 {\eta_{j}(\eta_{j}\!-\!1)} = {j(j+1}), \, j=2,3,4, 
\end{eqnarray}
\noindent
where
\begin{eqnarray}
\eta \equiv {{a}\over{\epsilon_{j}}} {d\epsilon_{j}\over{da}}
,\end{eqnarray}
\noindent
and where $a$ is the mean radius of the equipotential surface within the star,  $\epsilon_j$ is the tesseral 
harmonics of order $j$, then  $\rho(a)$ is the mass density at the distance {\it a} from the centre, and $\overline\rho(a)$ is the 
mean mass density within an 
equipotential of radius {\it a}. The following boundary conditions were applied: $\eta_{j}(0)$ =$j-2$, and 
$\left(d\eta_{j}\over{da}\right)_{\eta=0}$ = $-{3(j-1)\over{j+1}}{dD\over{da}}$, where $D={\rho(a)/{\overline\rho(a)}}$. 

The moment of inertia was integrated simultaneously using the  following equation:

\begin{eqnarray}
I = {8 \pi R^2\over{3}} \int_{0}^{R} \rho(r) r^4 dr, 
\end{eqnarray}

\noindent
where R is the radius of the configuration and $\rho(r)$ the local density. The radius of gyration was computed by 
using a simple  equation:

\begin{eqnarray}
        \beta =  \left(I\over{M R^2}\right)^{1/2}.
\end{eqnarray}
Equations 11 and 13 were integrated simultaneously
through a fifth-order Runge-Kutta method, with a tolerance level 
of 1$\times$10$^{-7}$.

\section {Tidal evolution of TZ For}

The eccentricities of the orbit of TZ For as measured by Andersen et al. (1991) and Gallene et al. (2016) 
are very similar (0.000 and 0.00002)  and so we can assume the orbit to be circular. 
 At this point, we can analyse two possible scenarios: the first one assuming
a high initial eccentricity and the other an initially circular system.
 To get the eccentricity {\it e}, $\Omega_1/\omega $, 
$\Omega_2/\omega$ and the orbital period as a function of time  Eqs. 1-4 were  integrated using 
as input the internal structure of  theoretical stellar models discussed in Section 2.1. A necessary condition 
for such integrations 
is knowing the boundary conditions of the differential equations. As explained in Claret (2011), 
such "observational"\ initial values are  unknown. However, we can introduce some initial test values for the 
eccentricity, orbital period, and angular velocities to study the tidal evolution.  Finally, using 
this procedure we obtain the evolution of these parameters as a function of time  to be  compared   
with their observed counterparts.

The rotational velocity measured by Gallene et al. (2016) for the secondary component, V$_2$sin~i = 45.7$\pm$1 km/s, 
is not very different from that measured by Andersen et al. (1991), namely: V$_2$sin~i  = 42$\pm$2 km/s. 
However, for the primary   the rotational velocities given by the authors  are  6.1$\pm$0.3 km/s and 4.0$\pm$1, 
respectively. 
Here, we adopt the most recent measurements by Gallene et al. (2016) to obtain the observational  
ratios $\Omega_1/\omega$ = 1.10$\pm$0.03 and $\Omega_2/\omega$ = 17.60$\pm$0.03. 

\subsection {Case of a highly eccentric orbit}

As we can verify by inspecting Eqs. 1-4, the evolution of the orbital parameters is strongly dependent on 
the relative radii of the components (eighth and sixth power). At  the present evolutionary state of both components 
of TZ For,  the relative radii contribute  only slightly to the tidal 
evolution of the system (around 0.07 and 0.03), respectively. Figure 2 shows the evolution of the orbital period  
as a function of time for the case  of an initial   P$ _{orb}$ = 80.1 days, $e_{initial}$ = 0.30 and 
$\Omega_1/\omega$ = $\Omega_2/\omega$ = 21.5.   
The pronounced peak in the period at log $\tau$  = 9.01 ($\tau$ represents the time in years) 
is due to the relative radius of the primary reaching its  maximum ($\approx$ 0.23), while the respective 
value for the secondary is  on the order of only 0.03. Such a peak is statistically very difficult to detect observationally 
due to the short time interval during which it occurs.  From this point on, the period remains practically constant 
and consistent, within the uncertainties of the stellar models and with its current observational counterpart. 

A very  useful  quantity in tidal evolution studies is the timescale (or critical time) for circularisation, 
 or the corresponding log g$_{cri}$,   since it works as a primary diagnostic tool. 
In the present case,  such critical values were computed until the initial eccentricity decayed to 0.368 $\times$ e$_{initial}$.   
In these calculations, it was assumed that the system is synchronized, namely, $\Omega_{1,2}$ = $\omega$ and e$_{initial}$ = 0.30. 
Figure 3  illustrates the  log g$_{cri}$ or equivalently the time scale as a function of the orbital period (black line). 
The primary (red square), at the age log $\tau$$\approx$ 9.01 for which the orbit becomes circularized,   
 is above  the critical circularisation curve. On the other hand, the secondary (open star) is below the mentioned curve. 
{ This is a preliminary indication of the only slight contribution of the secondary to the circularisation of the orbit. 
With regard to eccentricity resulting from the integrations of Eqs. 1-4, its behaviour is similar to that of the orbital 
period as shown in   Fig. 4, and it also decreases rapidly at  log $\tau$  $\approx$ 9.01.  On the other hand, several 
numerical tests reveal that, at least for the case of TZ For, the  times for the circularisation of the orbit for these MESA 
models do not depend strongly on 
the initial values of the eccentricity.

Concerning  the synchronisation  of the two components  the situation is more complex (Fig. 5). 
Due to the strong impact of the changes in the radius of the primary in the tidal evolution, this component reaches the 
synchronism at the same time of stabilisation of the orbital period and of circularisation of the orbit. The effect of 
the differential evolution is clearly noted since the secondary is relatively far from synchronism given that its angular 
velocity is approximately 17  times faster than that of the primary component  for the age derived for TZ For (log $\tau$ $\approx$ 9.053). 
Figure 6 shows the orbital period and the rotational velocity of each component as a function of log $\tau$. The respective 
theoretical values are also in good agreement with the observational data from Gallenne et al. (2016). The spikes  in the 
rotational velocities of the primary and secondary in the intervals of log $\tau$ $\approx$ 9.00-9.013 and 9.050-9.072, 
respectively, are a consequence of the rapid variations of the surface gravities of both stars (Fig. 6, lower left corner). 
All these results are in agreement with the observed eccentricity of the system,  the measured rotational velocities of 
both components, and  the orbital period  considering the intrinsic uncertainties of the evolutionary models (opacities, loss of mass, 
equation of state, etc.) and the uncertainties in the friction timescale (Section 3.3).

Within this scenario,  the nuclear evolution of the primary component  is the main activity responsible for the circularisation of 
the orbit  of the system, for the orbital period stabilisation,   and also for the primary   synchronisation at log $\tau$ $\approx$ 9.01.  
It remains synchronized until  the derived age of TZ For, since its relative radius decreases, thus contributing  little to its evolution by 
tides after the maximum value of the relative radius r$_1$. 
The case of the secondary component is 
somewhat different. Despite the small difference in mass, the secondary has a small relative radius at log $\tau$ $\approx$ 9.01
 while r$_1$ $\approx$ 0.23 and therefore the tidal forces are  still not sufficient to act on its synchronisation that will occur 
 later,  namely, at log $\tau$ $\approx$ 9.065.

\subsection {Case of  initial circular orbit}

The results of the tidal evolution for the case of an initial circular orbit with 
the following parameters:  initial orbital period P$_{orb}$ = 70.0 days, $e_{initial}$ = 0.00, and 
$\Omega_1$/$\omega$ = $\Omega_2$/$\omega$ = 18.2 can be seen in Figs. 7 and 8. Unlike the case with 
$e_{initial}$ = 0.30, the orbital period increases from 70 to 75 days  stabilising at log $\tau$ $\approx$ 9.01. 
On the other hand, Fig. 8 shows that the angular velocities of the two components present an aspect similar to that of Fig. 5, 
except for the shape of the peak around log $\tau$ $\approx$ 9.01. 
The synchronisation times for the two components  are approximately the same as those obtained for a very eccentric 
initial orbit. Taking into account the limitations of the stellar evolutionary models and the friction timescale, t$_F$, 
we can consider  that the theoretical levels of synchronism of both components; although two very different initial 
conditions have been used, the time evolution of the eccentricity and orbital period   are also consistent with their actual  
observational equivalents at the mean age determined for TZ For.

\begin{figure}
\includegraphics[height=8.cm,width=6cm,angle=-90]{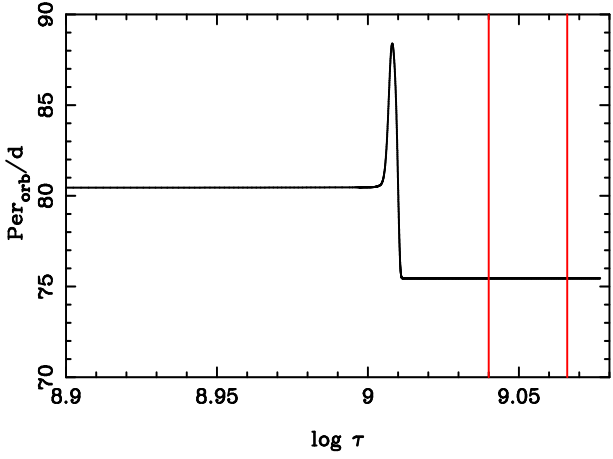}
\caption{  Evolution of the orbital period as a function of time.
The two vertical lines indicate the error bars for the mean age of the system. 
 Initial P$ _{orb}$ = 80.1 days, $e_{initial}$ = 0.30, and 
$\Omega_1/\omega$ = $\Omega_2/\omega$ = 21.5. }
\end{figure}

\begin{figure}
        \includegraphics[height=8cm,width=6cm,angle=-90]{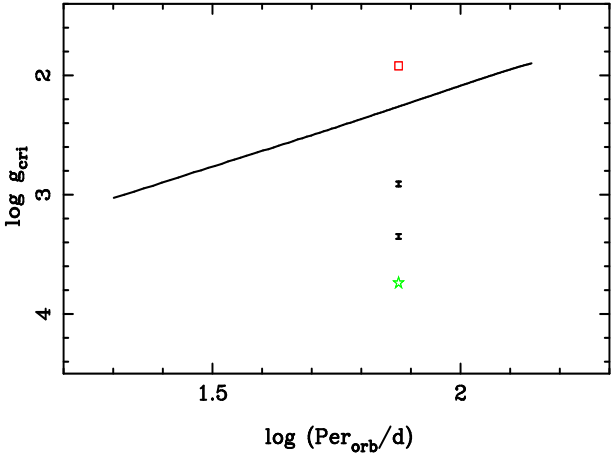}
        \caption{Critical values for circularisation as a function of the orbital period for the case of TZ For (continuous line).
                 The tiny black error bars represent the actual position of the TZ For, the square and the  filled star denotes 
                the primary and secondary component, respectively, at the time the system achieves  circularisation. 
 }
\end{figure}

\begin{figure}
        \includegraphics[height=8cm,width=6cm,angle=-90]{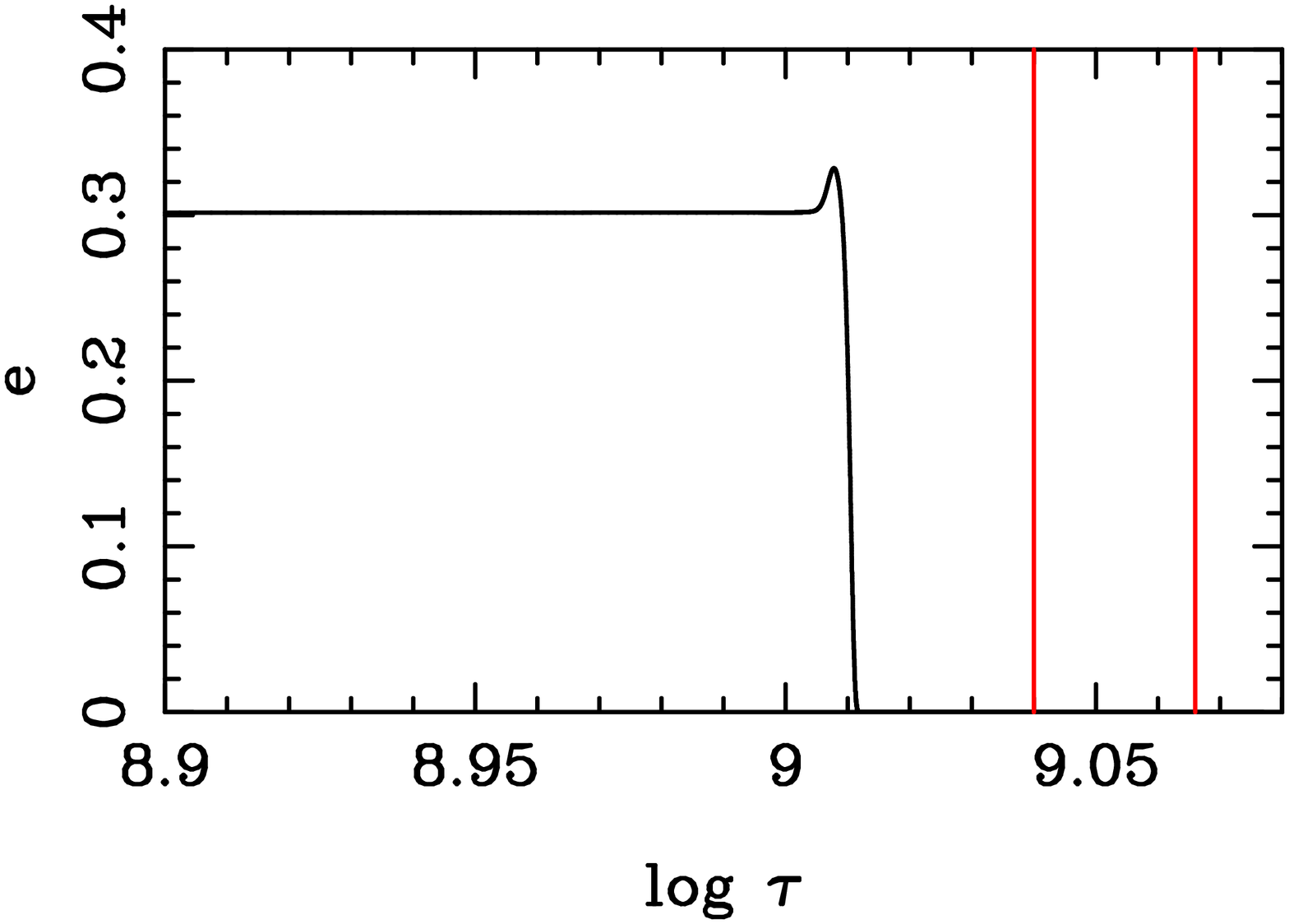}
        \caption{ Evolution of the eccentricity as a function of time.
                Same details as in Fig. 2. }
\end{figure}

\begin{figure}
        \includegraphics[height=8.cm,width=6cm,angle=-90]{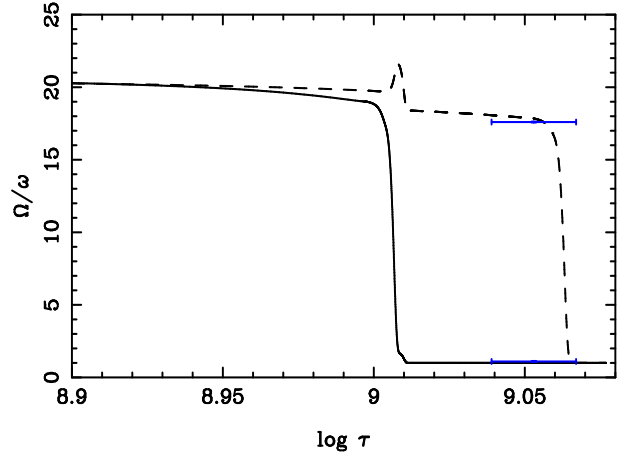}
        \caption{Evolution of the levels of synchronism as a function of time. The angular velocities of the two components 
were normalized to the mean angular orbital velocity. 
The continuous line represents the primary and the dashed one denotes the secondary.         
Same details as in Fig. 2.}
\end{figure}

\begin{figure}
        \includegraphics[height=8.cm,width=6cm,angle=-90]{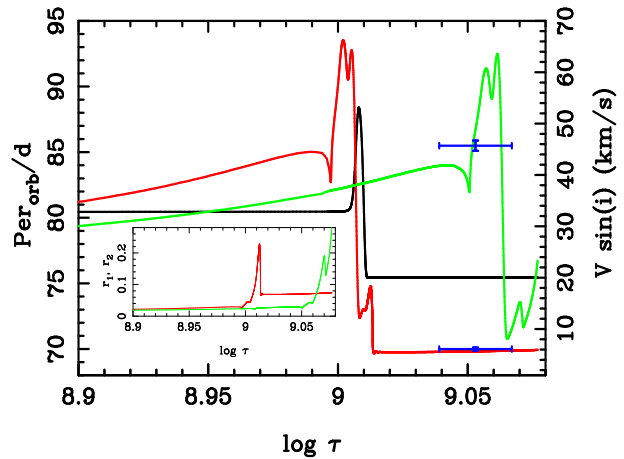}
        \caption{Time evolution of the orbital period (black line) and of the rotational velocities, both primary (red line)   
                and secondary (green line) as well as their respective  error bars (blue). The figure in the lower left corner 
                illustrates the variations in the relative radii of the primary and secondary. Same details as in Fig. 2.}
\end{figure}

\begin{figure}
        \includegraphics[height=8.cm,width=6cm,angle=-90]{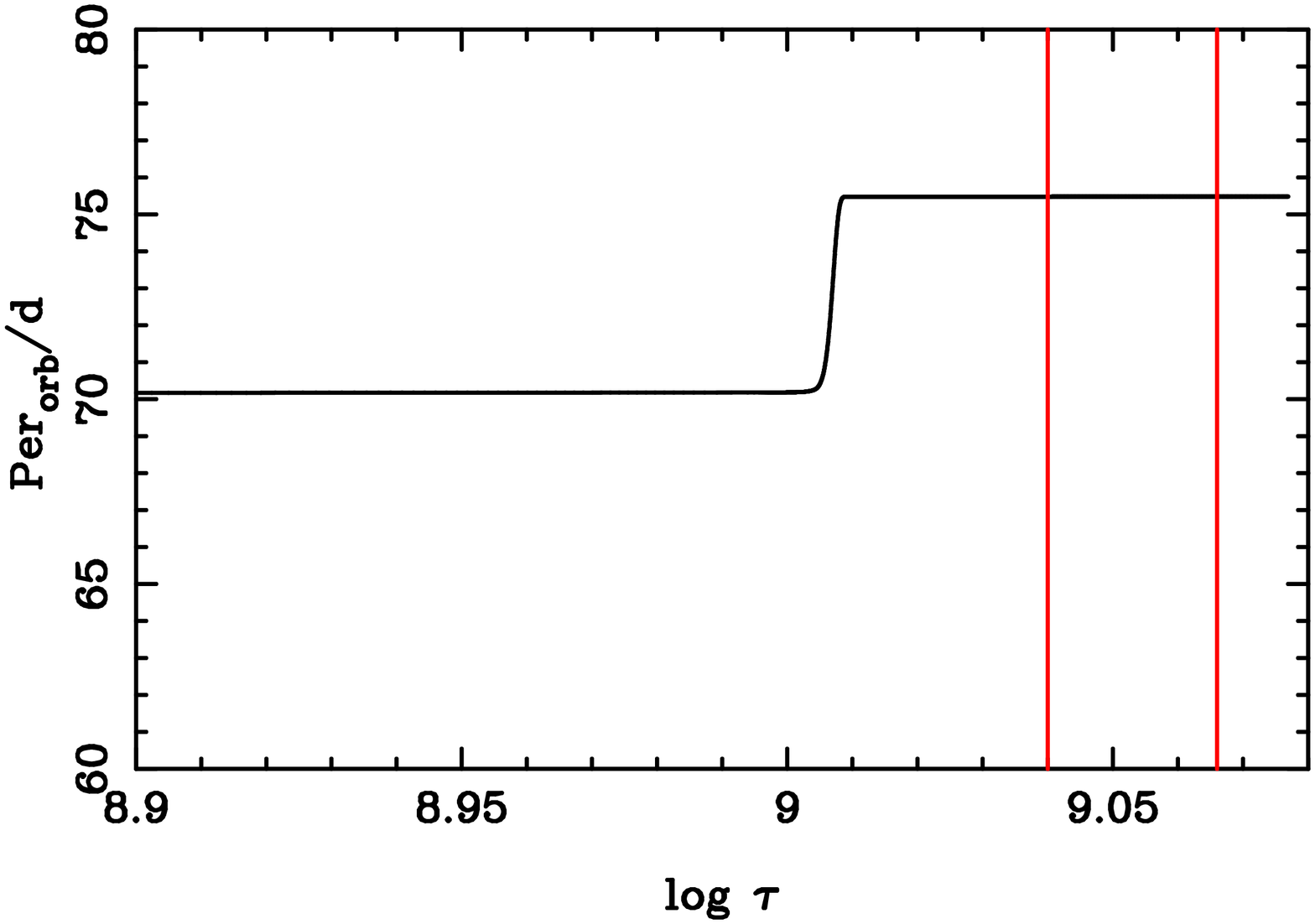}
        \caption{Evolution of the orbital period as a function of time.
                The two vertical lines indicate the error bars for the mean age of the system. 
                 Initial P$ _{orb}$ = 70.0 days, $e_{initial}$ = 0.00, and 
                $\Omega_1/\omega$ = $\Omega_2/\omega$ = 18.2.}
\end{figure}

\begin{figure}
        \includegraphics[height=8.cm,width=6cm,angle=-90]{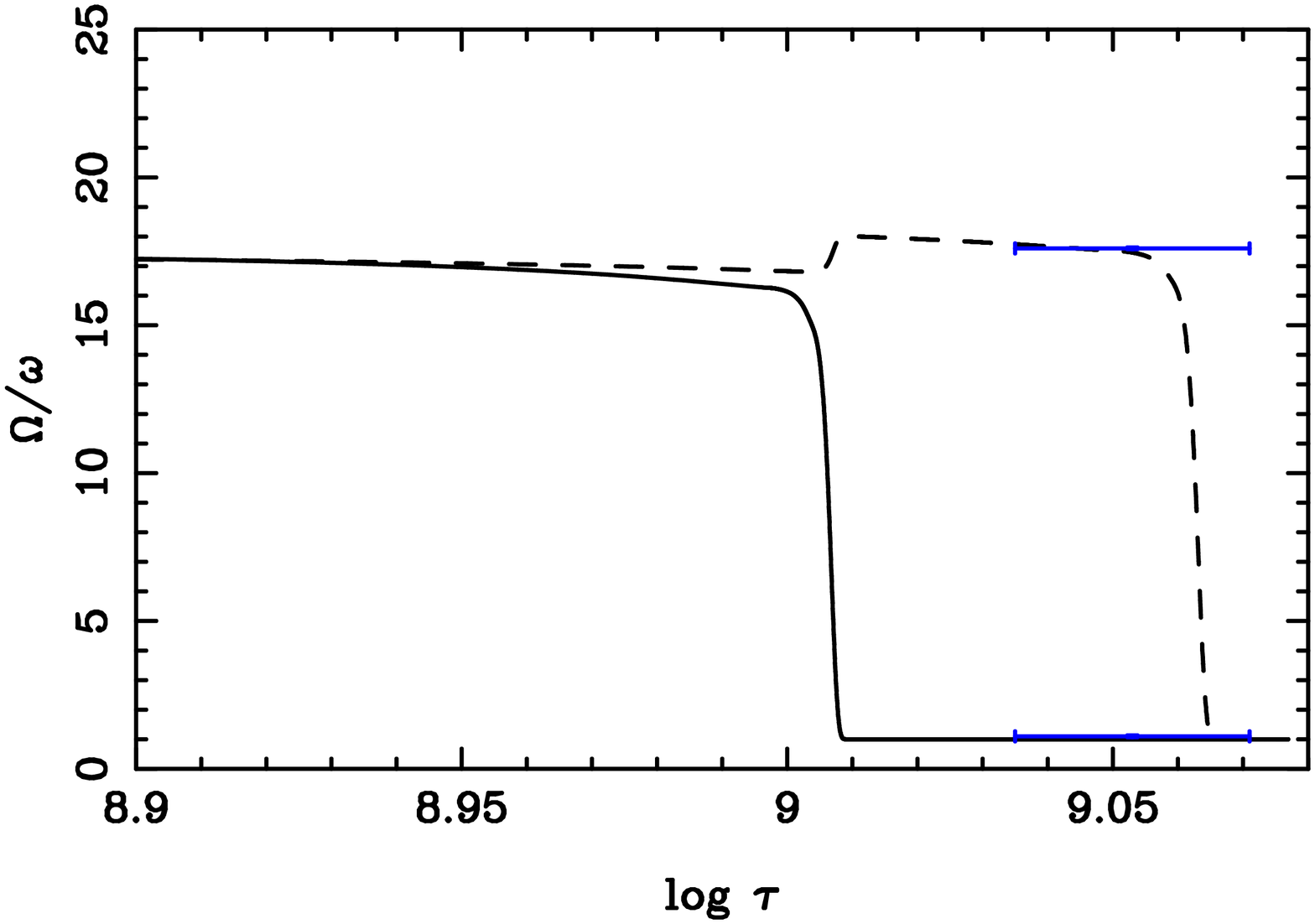}
        \caption{Evolution of the levels of synchronism as a function of time.
                The continuous line represents the primary and the dashed one denotes the secondary. Same details as in Fig. 7.}
\end{figure}

\begin{figure}
        \includegraphics[height=8.cm,width=6cm,angle=-90]{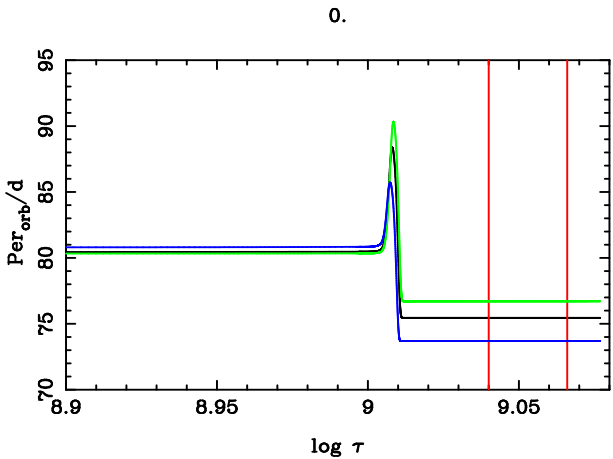}
        \caption{Effects on the orbital period due to uncertainties in the friction time scale t$_F$. 
The black line represents the evolution of the period as in Fig. 2 while the green line denotes the case [1.5$\times$t$_F$] 
and the blue one illustrates the case [0.5$\times$t$_F$]. Same initial conditions as in Fig. 2.  }
\end{figure}

\begin{figure}
        \includegraphics[height=8.cm,width=6cm,angle=-90]{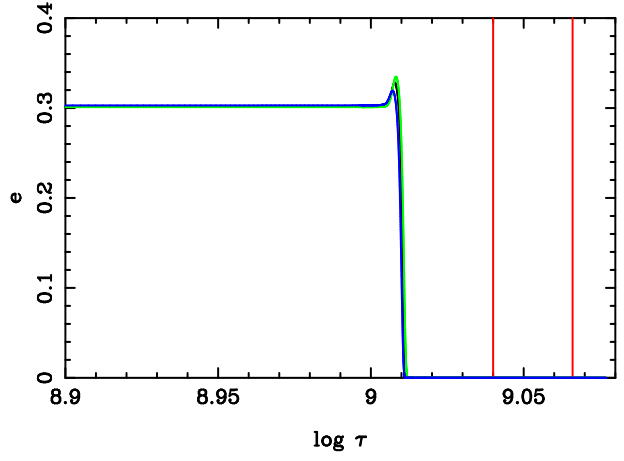}
        \caption{Effects on the eccentricity due to uncertainties in the friction time scale t$_F$. 
                The black line represents the evolution of the eccentricity as in Fig. 3 while the green line denotes the case 
                [1.5$\times$t$_F$] and the blue one illustrates the case [0.5$\times$t$_F$]. Same initial conditions as in Fig. 2.  }
\end{figure}

\begin{figure}
        \includegraphics[height=8.cm,width=6cm,angle=-90]{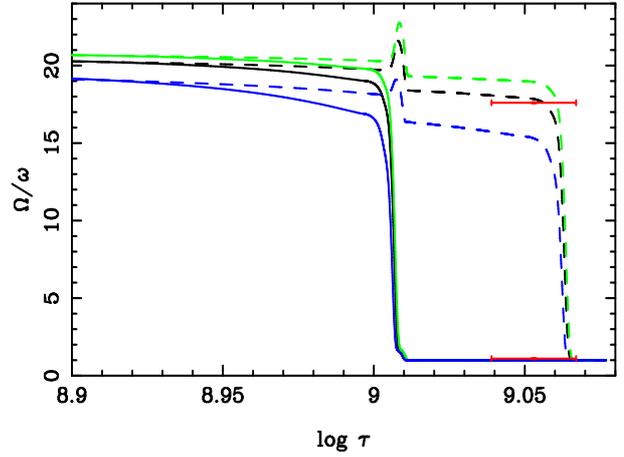}
        \caption{Effects  on the levels of synchronism of TZ For due to uncertainties in the friction time scale t$_F$. 
                The black line represents the evolution of angular velocities  as shown in Fig. 5, while the green line denotes the case of 
                [1.5$\times$t$_F$] and the blue one illustrates the case of [0.5$\times$t$_F$]. The continuous lines denote the primary component 
                and the dashed ones the secondary. The error bars are represented in red.  Same initial conditions as in Fig. 2.  }
\end{figure}

\begin{figure}
        \includegraphics[height=8.cm,width=6cm,angle=-90]{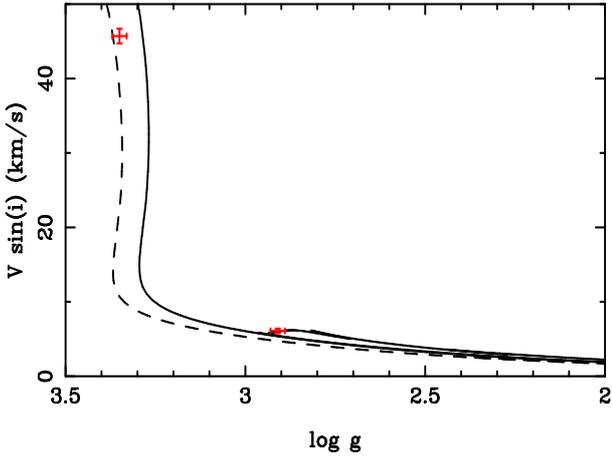}
        \caption{Rotational velocities for TZ For for the case of  solid body rotation. 
 Continuous line represents the primary and the dashed one denotes the secondary.                 
                 The calculations were performed following a revised version of GRANADA code by Claret (1999). The derived mean age 
                 of the system is 1.23$\pm$0.05 Gyears. }
\end{figure}

\begin{figure}
        \includegraphics[height=8.cm,width=6cm,angle=-90]{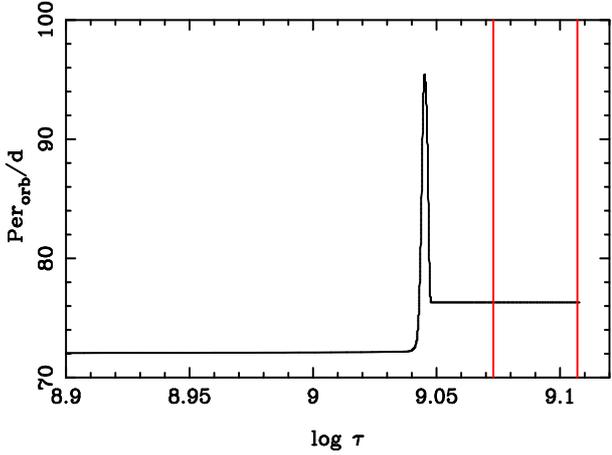}
        \caption{Evolution of the orbital period as a function of time.  Initial P$ _{orb}$ = 72.03 days, $e_{initial}$ = 0.30, and 
                $\Omega_1/\omega$ = $\Omega_2/\omega$ = 56.0. The initial conditions for the angular velocities of the two components 
                are those of the GRANADA models with rotation near the ZAMS.  The two vertical lines indicate the error bars for the 
                mean age of the system. }
\end{figure}

\begin{figure}
        \includegraphics[height=8.cm,width=6cm,angle=-90]{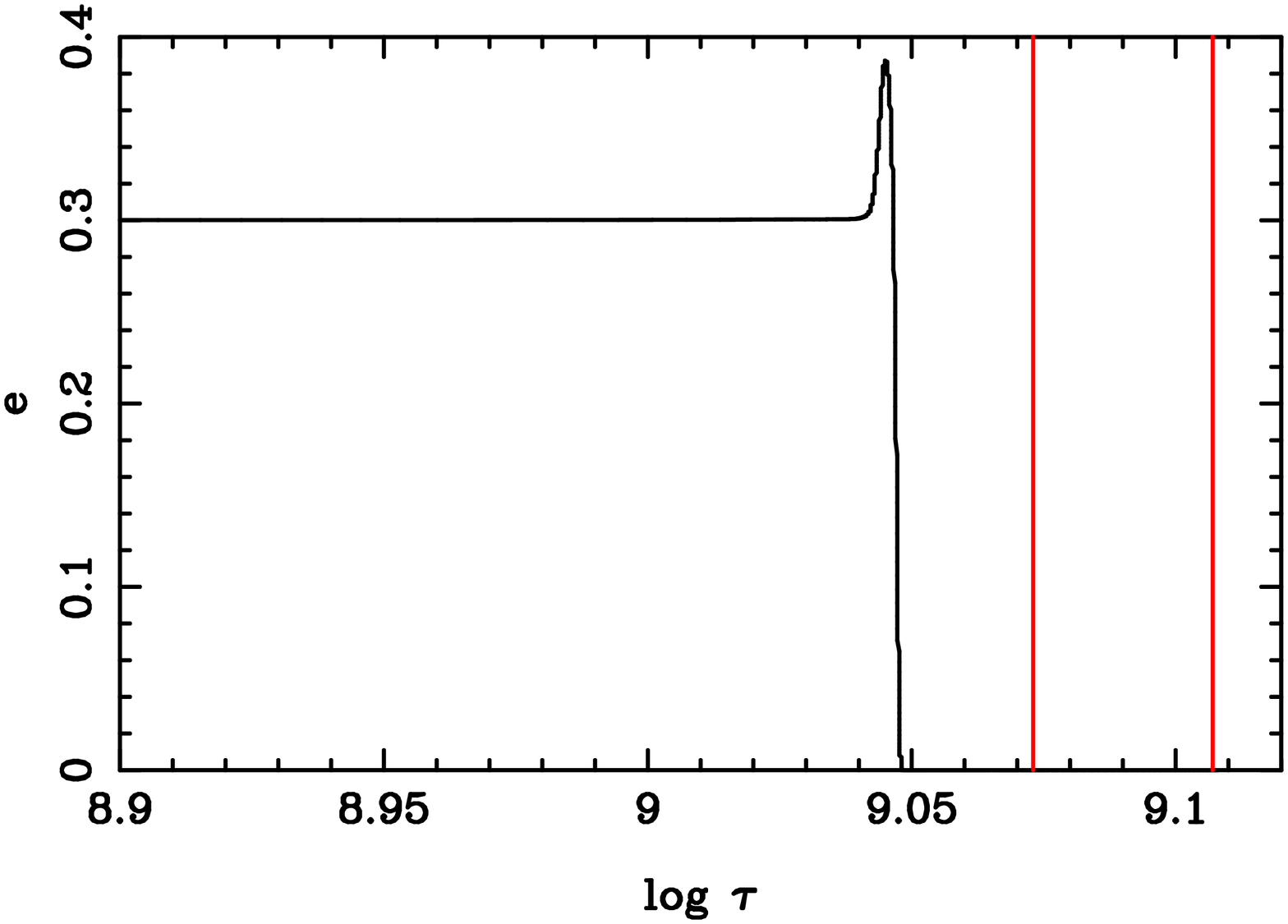}
        \caption{Evolution of the eccentricity as a function of time. 
        Same details as in Fig. 13}
\end{figure}

\begin{figure}
        \includegraphics[height=8.cm,width=6cm,angle=-90]{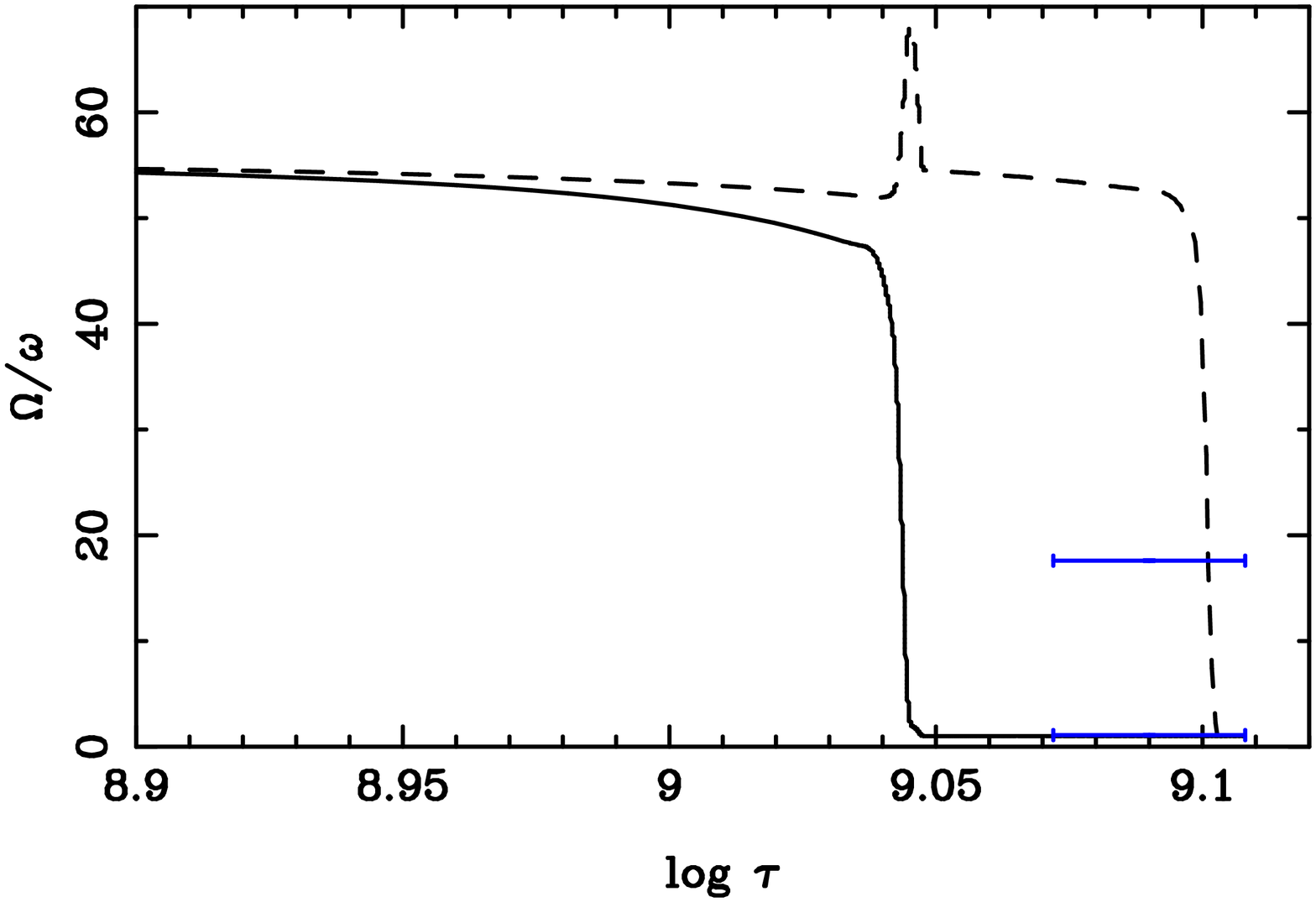}
        \caption{Evolution of the levels of synchronism as a function of time. 
        The continuous line represents the primary and the dashed one denotes the secondary.   Same details as in Fig. 13.}
\end{figure}

\begin{figure}
        \includegraphics[height=8.cm,width=6cm,angle=-90]{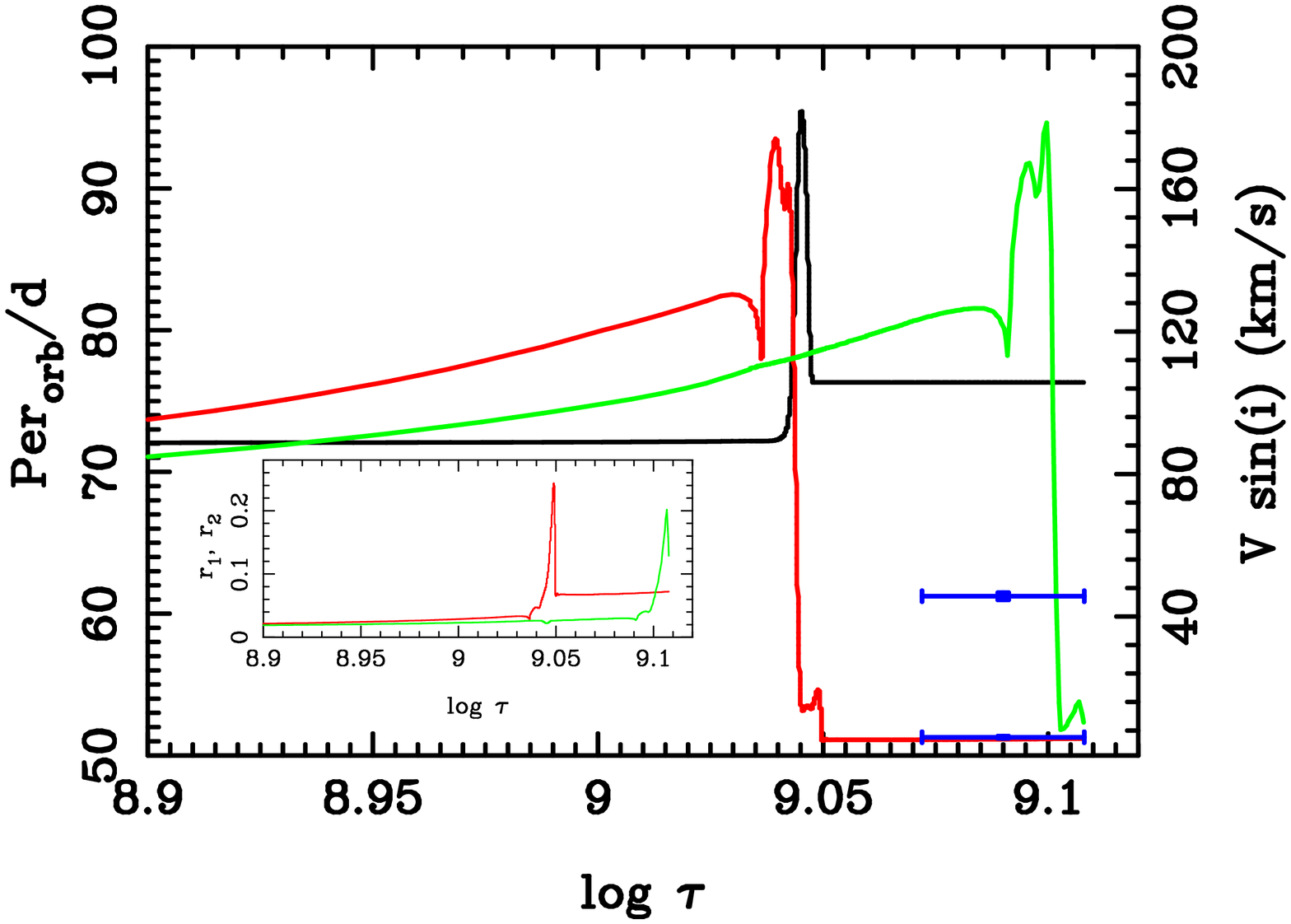}
        \caption{Rotational velocities, according to the integrations of Eqs. 1-4,   as a function of time   for the 
                rotating models shown in Fig. 12.  
                Continuous red line represents the primary and the green one denotes the secondary.    The two blue  
                lines indicate the error bars for the mean age of the system. 
                Lower-left corner illustrates the variations in the relative radii of the primary and secondary.     
                Same initial conditions as in Fig. 13. }
\end{figure}

\subsection {Influence of the uncertainties of t$_F$ on the tidal evolution of TZ For}

As we have seen, an acceptable agreement has been achieved between the evolutionary models and the absolute 
dimensions of TZ For with an error in age   $<$  5\%.  
In the previous section, we show that the tidal theory in the weak friction approach is capable of 
explaining
the current orbital period, the behavior of the eccentricity, and the
synchronization levels of the two components.
         However, one of the weakest points  in Eqs. 1-4  is the friction time-scale,  t$_F$. 
          In this section, we try to estimate  
         the influence of friction time uncertainties on the tidal evolution of TZ For. For this   purpose,  
         we adopted a very simple scheme: 
         we assume that the errors in t$_F$ are  of $\pm$ 50\%. 
As before, we integrate the aforementioned equations changing t$_F$ for the case P$ _{initial}$ = 80.1 days, 
$e_{initial}$ = 0.30, and 
$\Omega_1/\omega$ = $\Omega_2/\omega$ = 21.5 that was taken as the reference.

Figure 9 illustrates the influence of considering errors in t$_F$ of $\pm$ 50\% in the evolution of the orbital 
period where we have 
taken the results shown in Fig. 2 (black line)  as a reference. The differences  $\Delta$P/P in the stabilisation 
zone of  the  period 
are respectively 1.4\% for the case of [1.5$\times$t$_F$] and 2.5\% for the case of [0.5$\times$t$_F$]. Such differences 
are acceptable given 
the mentioned intrinsic uncertainties in the calculation of  the theoretical stellar models.  
In relation to the evolution of the eccentricity (Fig. 10) the differences in the circularisation times are almost 
indistinguishable 
for the three configurations. 

The evolution of normalized angular velocities is more complex than the case of eccentricity (see Fig. 11) since 
they present the greatest  
differences. For example, the values of the normalized rotational velocities  of the secondary, considering the  
mean age of TZ For, differ 
up to  6\% and 14\% for [1.5$\times$ t$_F$] and [0.5$\times$t$_F$], respectively.

We also note   that for a 50\% larger value of $t_F$,  the normalized angular velocities are approximately equal to 
or larger than those of the 
reference model; the opposite occurs  reducing by a half  t$_F$. It is also remarkable that in both cases, 
the synchronization times 
for the two  components converge to log $\tau$  $\approx$ 9.01 and 9.065, respectively,  even though the normalised 
angular velocities are 
different before this time. We can consider that despite assuming uncertainties in t$_F$ on the order of 50\% 
and also considering the 
intrinsic uncertainties in the evolutionary models, the synchronization levels of the two components, the 
circularisation time, and the 
stabilization time of the orbital period are compatible with their observational counterparts.

\subsection {Another hypothetical scenario: Estimation of the role of  rotating models}

The fact that the current orbital period of TZ For is long implies some peculiarities. The behaviour of the orbital period  
(Figs. 2, 7 and 9)  shows    very accentuated changes in this variable during only a short interval of time (log $\tau$ $\approx$ 9.01). 
Such rapid changes   practically govern the  tidal evolution of the system due to the impact of the changes in the relative radii, k$_2$,  
and the moments of inertia, mainly in the case of the primary component.  Outside this  short interval of time the orbital period remains 
practically constant. Of course, there are many more theoretical scenarios that we have on hand to explain the current state of TZ For's orbital elements, beyond 
those presented in Sections 3.1, 3.2, and 3.3. Here, we try to address one of  these, which is focused on rotation.  In fact, a  rotating stellar model evolves differently from a standard one. The  differences will be larger if additional phenomena, such as turbulent diffusion, core overshooting, or rotational-mixing, are taken into account.  The evolution of a   rotating model 
will differ from the standard one in terms of luminosity, effective temperatures, lifetimes of Hydrogen and He burning, and changes in the surface 
chemical abundances,   as well as  its internal structure ( k$_2$ and  
 $\beta$). Differently from the results presented in the previous sections, 
we  investigate here  the role of rotating models in the evolution of the orbital elements of  TZ For. For this purpose, we introduced rotation in the GRANADA code using the method  by  Kippenhahn \& Thomas (1970),  improved by Endal \& Sofia (1976, 1978) 
and adapted in this code by Claret (1999, 2022). Such a code is designed to treat systems disturbed by rotation and by tidal forces. To 
take into account only the effects of rotation, we assumed the mass ratio to be $q$ = 0. We adopted the  solid body and overall conservation of angular momentum  approach for simplicity. 
 However, such rotating models present some limitations:  
        by definition, in such  models, the angular velocity is the same throughout the model. Another important simplification is that rotational 
        mixing has not been considered (only core overshooting is taken into account). These  simplifications  limit the scope of our conclusions, 
        but they allow us to estimate the effects of the  departure from spherical geometry  on $k_2$ and on the radius of gyration which are key 
        for the tidal evolution calculations.

 In searching for the best solution, we used the same method and the statistics $\chi^2$ described above for the MESA models, also  taking 
        into account the grids of  initial angular velocities.  
The input physics  for the best solution for the rotating models was:  X = 0.711, Z = 0.015, $\alpha_{MLT1}$ = 1.82, $\alpha_{MLT2}$ = 1.82, 
$\alpha_{ov1}$ = 0.23, and $\alpha_{ov2}$ = 0.21. The adopted opacities are based on the mixture of   Asplund et al. (2009). We note  that 
convective core overshooting in the GRANADA code is simulated 
adopting a step-function, characterized by the parameter $\alpha_{ov}$ whose relation with f$_{ov}$ is given by $\alpha_{ov}$/f$_{ov}$ = 11.36$\pm$0.22 
following Claret\&Torres (2017). The initial angular  velocities were $\Omega_1$ = 5.2$\times$10$^{-5}$ s$^{-1}$ and $\Omega_2$ =  4.9$\times$10$^{-5}$ s$^{-1}$. 
The rotacional velocities of both components reaching the ZAMS (zero age main sequence) are 84.0 and 77.0 km/s, respectively. 
These values  are compatible with the observational data for DLEBS  with similar masses, as for example, the case of V1647 Sgr, whose 
observational counterparts are, respectively, 80$\pm$5 km/s and 77$\pm$5 km/s (Torres et al. (2010), Table 2).  The derived common age 
was 1.23$\pm$0.05 Gyears. The difference between the ages of TZ For as derived using the MESA and GRANADA codes are due mainly to the 
different nuclear network adopted in both codes (see Claret (2004), Section 2). Furthermore, the introduction of rotation also influences 
the determination of the ages.  
For  details on the implementation of rotation in GRANADA models, we refer to Appendix A.

The results of these calculations can be seen in Fig. 12, where the rotational velocities are represented as a function of log g. As  it can be 
inspected in such a figure,  the theoretical values of the rotational velocities  are in good agreement with their observational counterparts. 
In such a figure, it can be verified that both components of TZ For are in the same evolutionary stages as in the case shown in Fig. 1. In fact, 
the corresponding HR Diagram for such models corroborates this point.  This similarity indicates that, although the present evolutionary models 
were computed using a  different code and input physics from that discussed in Subsection 2.1, the intercomparison leads to  similar results 
concerning the evolutionary status of TZ For and also  the respective internal structure (k$_2$ and moment of inertia). This  shows the 
consistency between the two stellar evolution codes used in the present study, despite the differences in the respective nuclear networks and 
the implementation of rotation in the GRANADA models. 

In order investigate the effects of  the intrinsec stellar rotation,    we  proceeded to calculate the evolution by tides using the rotating 
GRANADA models. As a novelty, we did not use trial values for the rotational velocities (as we had done in the case of the MESA models). Here, 
we adopt the initial $\Omega_i$ derived from GRANADA models with rotation (Fig. 12) as initial conditions.
The similarities between Figs. 2, 4, and 5 (MESA) and their corresponding ones using the rotating GRANADA models (Figs. 13, 14, and 15) are 
notorious, except for the timescale and the initial orbital period adopted, which is as expected. This fact confirms again that the internal structures 
of both models are very similar. On the other hand, in Fig. 16 we can see that such a similarity also extends to the rotational velocities. 
Comparing the rotational velocities shown in Figs. 6 and 16, we note that both are morphologically similar. However, they  differ regarding 
the values being the ones shown in Fig. 16 systematically larger.  The reason for this is that in Fig. 16, the adopted initial period is 
smaller than its counterpart in Fig. 6. In addition,  the corresponding angular  velocities   were computed with initial values of  $\Omega_i/\omega$ 
larger than in the case shown in Fig. 6 (56.0 and 21.5, respectively). In fact, the maximum value of the  relative radius of the primary in Fig. 16 is 
larger than that shown in Fig. 6 (0.25 and 0.23, respectively). This  implies, for example,  that the tidal interaction is about 2.0 times stronger 
in the case of Fig. 16 during such  maxima. 

\subsection {Summary}

We studied the nuclear and tidal evolution of the TZ For system using evolutionary tracks computed with two different 
codes. Such models  reproduce, within the observational  errors, the absolute dimensions of the system with a tolerance 
of 5\% in the common age.

Regarding tidal evolution, we integrated the equations of Hut (1981) in the cases of high (0.30) and low (0.0) 
initial eccentricities. Good agreement has been found between the observed values of eccentricity, orbital period, and 
synchronism levels with their theoretical counterparts. The influence of friction time in such calculations was also studied and we have concluded that its influence affects the synchronization levels of both components is greater, 
although we consider that they are acceptable given the intrinsic uncertainties in t$_{F}$ and in the stellar evolutionary 
models.

On the other hand, we computed rotating evolutionary tracks with the code GRANADA. Such models, which also 
reproduce the absolute dimensions of TZ For, were used to estimate the influence of rotating models on tidal 
evolution. For this purpose, we used the theoretical  initial angular velocities of the two components as the initial 
conditions for the integrations of Eqs. 1-4. These values do not replace the observational values for which 
we do not have reliable data so far, but we believe it is a small step forward. As in the previous cases, we found a good agreement between the observed orbital elements and their corresponding  theoretical values. 

{}

\begin{acknowledgements} 
 I thank an anonymous referee for his/her careful reading of the original version as well as for 
        his/her comments and suggestions. 
The Spanish MEC (ESP2017-87676-C5-2-R,  PID2019-107061GB-C64, and  
PID2019-109522GB-C52) is gratefully acknowledged for its 
support during the development of this work. A.C.  
acknowledges financial support from the State Agency for 
Research of the Spanish MCIU through the “Center of 
Excellence Severo Ochoa” award for the Instituto de 
Astrofísica de Andalucía (SEV-2017-0709).  This research has made use of the SIMBAD database, 
operated at the CDS, Strasbourg, France, and of NASA’s Astrophysics Data System
Abstract Service.

\end{acknowledgements}

\begin{appendix} 
        
        \section{Brief description of the numerical method for computing stellar rotating  models}

 The  formalism by Kippenhahn \& Thomas (1970) has been
used to simulate  the effects of the stellar rotation on the internal density
concentration as well as on the evolution of the rotational velocity. The mathematical basis of this method can be found in 
Kippenhahn \& Thomas (1970) and improved by Endal \& Sofia (1976, 1978). A  star distorted  by rotation and by tides  
is represented by a Roche's critical surface. The suitable potential is written in terms of massratio, relative distance 
and  angular velocities. This potential presents some advantages, as for example: if the only interest is to investigate 
the effects of  stellar rotation, then  the mass ratio ought to be $q$ = 0.0.

To simulate rotating stars we restricted the calculation to first order theory and  the total potential  
can be written as given by Kopal (1959): 

\begin{eqnarray}
        {\psi} = {GM_{\psi}\over{r^2}} + {1\over{2}}\Omega^2r^2 sin^2\theta - \nonumber \\
        {4\pi G\over{5r^3}}P_2(cos\theta)\int_{0}^{a}\rho {d\over{da'}}\left(a'^5f_2\right)da'
,\end{eqnarray}

\noindent
where

\begin{eqnarray}
        r = a\left(1 - f_2 P_2(cos\theta)\right)
,\end{eqnarray}

\begin{eqnarray}
        f_2 = {5\Omega^2a^3\over{3GM_{\psi}(2+\eta_2)}}.  
\end{eqnarray}

In the above equations $\Omega$ is the angular velocity, P$_2$(cos$\theta$) is the second Legendre polynomial, 
$a$ the radius of the level surface, $\eta_2$ is the solution of  the  Radau's equation (Eq. 11), and 
the remaining symbols retain their usual meaning.

In this framework a generic function F(r, $\theta$, $\phi$) has its counterpart according to:
\begin{eqnarray}
 {\overline F}= S^{-1}_{\psi}\int {{FdS}}, 
\end{eqnarray}

\noindent
where dS is the element of surface for constant values of  $\psi$.
On the other hand,  the effective gravity is given 
 by:

\begin{eqnarray}
 {g} = {d\psi\over{dn}} 
,\end{eqnarray} 
 
\noindent
where {\it dn} is the distance between two neighbouring surfaces. 

The corresponding volume of the configuration is thus given by:

\begin{eqnarray}
V_{\psi} = {4\pi\over {3}} r_{\psi}^{3}.
\end{eqnarray}

The usual differential equations of the stellar structure are  changed  
to:

\begin{eqnarray}
{\partial r_{\psi}\over\partial M_{\psi}} = {1\over{4\pi\rho r_{\psi}
                ^2}}
,\end{eqnarray}  

\begin{eqnarray}
 {\partial P_{\psi}\over{\partial M_{\psi}}} = -{GM_{\psi}\over {4\pi
                r_{\psi}^4}}f_{P}
,\end{eqnarray}

\begin{eqnarray}
 {\partial L_{\psi}\over {\partial M_{\psi}}} = {\epsilon} -   {\partial E\over \partial t} - P{{\partial{1\over{\rho}}}\over{\partial t}}
,\end{eqnarray}          

\begin{eqnarray}
{\partial ln T_{\psi}\over{\partial ln P_{\psi}}} =
{3\kappa L_{\psi}P_{\psi} f_{T}\over{16\pi acGM_{\psi}T^4_{\psi} f_{P}}}. 
.\end{eqnarray}

In the above equations, $M_{\psi}$, $L_{\psi}$ are the mass and luminosity enclosed by a constant equipotential, 
$\epsilon$ is the  nuclear energy generation rate per unit mass, $E$ is the internal energy  per unit of mass, 
T and P the temperature and pressure, $\rho$ the density, $a$ the radiation pressure constant (not to be confused 
with mean radius of an equipotential),  and $c$ the velocity of light in vacuum. In principle, when using these 
approximations, it would be necessary to solve the Poisson equation simultaneously. We note that if we choose 
the potential as the solution  corresponding to the Roche model, such a condition is not necessary.

After some algebra the Schwarzschild criterion  can be written as: 

\begin{eqnarray}
{\partial ln T_{\psi}\over{\partial ln P_{\psi}}} = min [
\bigtriangledown_{ad},
{\bigtriangledown_{rad}}{f_{T}\over {f_P}}], 
\end{eqnarray}
        
\noindent
where $\bigtriangledown_{ad}$ and $\bigtriangledown_{rad}$ are the "spherical"$\emph{}$
adiabatic and radiative gradients. The variables f$_P$ and f$_T$ are computed 
following the equations: 

\begin{eqnarray}
 {f_P} = {4\pi r_{\psi}^4}{1\over {GM_{\psi}S_{\psi}\overline {g^{-1}}}}
,\end{eqnarray}

and

\begin{eqnarray}
 f_{T} = {\left(4\pi r_{\psi}^{2}\over S_{\psi}\right)}^2 {1\over
        {\overline g
                \overline {g^{-1}}}}.
\end{eqnarray}

The functions $f_P$ and $f_T$ depend on the shape of the equipotential surfaces and for $f_P$ and $f_T$ = 1.0 that we 
recover the spherical model. 
To simulate rotating models  we have used an updated version of the GRANADA code described in Claret in 1999 (and revised in 2022). Inspecting Eqs. A1-A3, we note that Eq. 11 should be integrated  simultaneously to obtain 
$\eta_2$  at each point of the model to solve the modified differential equations of stellar structure. 
This is also necessary for the calculation of average local gravity and its inverse.
Regarding the numerical method used to calculate the rotating models   we introduced the following procedure 
to take into account these requirements:  the
first model is computed without rotation. It will give, through
integrations, the values of $\eta_2$  and therefore of the integral in Eq. A1 for each point
to be used in a second model which includes rotation. For this
second model (and all the following ones), the values of the integral
and of $\eta_2$  will include the rotational effects. The model $n$ will be
computed considering the values of the integral and of $\eta_2$ as
derived from the model $n-1$.  Provided that neighboring models
have similar structures (small steps in time) the method guarantees 
a good accuracy. Finally, to compute $f_P$ and $f_T$, we need to use a relationship between $r_{\psi}$ 
and the mean radius of a given equipotential $a$. Such variables are connected by the following equation 
which is solved by iteration:

\begin{eqnarray}
        r_{\psi}^3 \approx a^3\left( 1 + {3\over{5}}f_2^2 - {2\over{35}}f_2^3\right). 
\end{eqnarray} 

For more detailed information on the implementation of rotation in the GRANADA code, we refer to Claret (1999).

\end{appendix}

\end{document}